\documentclass{aa}
\usepackage{graphicx}
\begin{document}
   \title{Early optical and millimeter observations of GRB 030226 afterglow} 

   \author{S.B. Pandey\inst{1}, R. Sagar\inst{1,2}, G.C. Anupama\inst{2}, D. Bhattacharya\inst{3}, 
   D.K. Sahu\inst{2,4}, A.J. Castro-Tirado\inst{5} and M. Bremer\inst{6}}

   \institute{ 
   State Observatory, Manora Peak, Nainital -- 263 129, Uttaranchal, India
   \and 
   Indian Institute of Astrophysics, Bangalore -- 560 034, India
   \and 
   Raman Research Institute, Bangalore -- 560 080, India
   \and
   Center for Research \& Education in Science \& Technology, Hosakote, Bangalore -- 562 114, India
   \and
   Instituto de Astrof\'isica de Andaluc\'ia, P.O. Box 03004, E-18080, Granada, Spain
   \and  
   IRAM, 300 rue de la Piscine, Dom. Universitaire 38406 Saint Martin d'Heres. France
}

   \authorrunning{S. B. Pandey et al.}
   \titlerunning{Early optical and millimeter observations of GRB 030226 afterglow}
   \offprints{S. B. Pandey, \\
           \email{shashi@upso.ernet.in}
            }
   \date{Received ------ /accepted ---------}

  \abstract{
The CCD magnitudes in Johnson $UBV$ and Cousins $RI$ photometric passbands
for the afterglow of the long duration GRB 030226 are presented. Upper 
limits of a few mJy to millimeter wave emission at the location of optical
are obtained over the first two weeks. The optical data presented here, 
in combination with other published data on this afterglow, show an early 
$R$ band flux decay slope of 0.77$\pm$0.04, steepening to 2.05$\pm$0.04 
about 0.65$\pm$0.03 day after the burst. Interpreted as the ``jet break'', 
this indicates a half opening angle of $\sim 3.2$~degree for the initial 
ejection, for an assumed ambient density of $\sim 1\;{\rm cm}^{-3}$.  
Broadband spectra show no appreciable evolution during the observations, 
and indicate the presence of synchrotron cooling frequency $\nu_c$ near 
the upper edge of the optical band. From the broadband spectra we derive an 
electron energy distribution index $p = 2.07\pm0.06$ and an intrinsic 
extinction $E(B - V)\sim0.17$. Millimeter upper limits are consistent with
these derived parameters.

} 

\maketitle
\section{Introduction}
Multi-wavelength observations of an afterglow of a Gamma Ray Burst (GRB),
an event which emits $\sim 10^{51} - 10^{53}$ ergs of energy in a few
seconds, provide valuable information about its progenitor (M\'{e}sz\'{a}ros, 2002).
Optical observations occupy the central place in such studies as they provide
information about distance as well as isotropic/non-isotropic nature of the 
emission (e.g. Sagar 2002). In this paper we present the early optical and 
millimeter-wave observations of the long duration ($>$ 100 sec) GRB 030226 
($\equiv$H10893).
It was detected by HETE FREGATE, WXM, and soft $X-$ray camera (SXC) 
instruments at $t_0 = 03^h 46^m 31.^s99$ UT on 26th Feb 2003 
(Suzuki et al. 2003).  Fox, Chen \& Price (2003) discovered its optical
afterglow (OA) within the HETE error circle at $\alpha = 11^h 33^m 04.^s9,
\delta = +25^{\circ} 53^{'} 55.^{''}6$ (J2000) with an uncertainty of $0^{''}.5$
in each coordinate (Price, Fox \& Chen 2003). $X-$ ray afterglow of the GRB 
030226 was discovered by Chandra $X-$ray observatory $\sim$ 37.1 hours after 
the burst at the OT position (Pedersen et al. 2003). Spectroscopic observations 
performed by Ando et al. (2003) on 2003 Feb 26.31 UT yield a value of redshift 
z $\sim$ 2.1, which was later refined to 1.99. VLT spectroscopic observations 
indicate z = 1.986 $\pm$0.001 or larger using FeII, AlII and CVI 
absorption lines (Greiner et al. 2003a). Based on Keck II high resolution spectra, 
Price et al. (2003) calculated the redshift systems at z = 1.043,1.948 and 1.961. 
Chornock \& Fillippenko (2003), using low resolution spectrograph on Keck I, 
deduce that the highest redshift system is associated with the GRB afterglow.

Dai \& Wu (2003) modeled the GRB 030226 afterglow light curve and concluded
that a discontinuous jump in the ambient density is indicated by the data.
However this work made use of the preliminary magnitude estimates communicated
in the GRB Coordinates Network (GCN), which show a large scatter.  This is
most probably due to the use of inhomogeneous photometric calibrations: those
given by Garnavich, von Braun \& Stanek (2003) and Henden (2003) differ by
0.06 and 0.01 mag in $R_c$ for two important comparison stars in the field,
designated $A$ and $B$ respectively. 
For reliable determination of parameters from the light curve of the
OA, consistent and secure photometric calibrations are needed. We provide this
by imaging PG0918+029 standard region of Landolt (1992) along with GRB 030226 
field.

\section {Observations and data reductions}  

The millimeter wave and broad band optical observations of the GRB 030226
afterglow are described in the following subsections.

\subsection{Millimeter Wave Observations }

The GRB 030226 afterglow was observed with the Plateau de Bure Interferometer 
(Guilloteau et al. 1992) in a six-antenna extended configuration on the
dates listed in Table 1. Flux calibration includes a correction for atmospheric 
decorrelation which has been determined with a UV plane point source fit of 
the phase calibration quasar 1156+295. The carbon star MWC349 has been used 
as a primary flux calibrator due to its well-known millimeter spectral
properties (see e.g. Schwarz 1980). There is no detection of millimeter wave
emission at the location of OA during two weeks of observations after the
burst (see Table 1).

\normalsize
\begin{table}
 {\bf Table 1.}~Millimeter wave observations of the GRB 030226 afterglow. 

\begin{center}
\scriptsize
\begin{tabular}{cccll} \hline 
Start & end & center & frequency & Flux  \\
time&time&time&(GHz)&center (mJy)  \\   \hline 
Feb 27.866&28.045&27.956&85.283&+0.54$\pm$0.31 \\ 
Feb 27.866&28.045&27.956&220.311&$-$0.90$\pm$1.96 \\ 
Mar 02.107&02.291&02.199&89.199&$-$0.50$\pm$0.76 \\
Mar 08.178&08.233&08.206&91.995&$-$1.01$\pm$0.70 \\
Mar 08.178&08.233&08.206&232.032&$-$0.83$\pm$3.45 \\
Mar 10.166&10.197&10.182&90.782&$-$0.43$\pm$0.84 \\
Mar 10.166&10.197&10.182&221.903&$-$4.00$\pm$3.86 \\
Mar 11.908&12.014&11.961&114.676&+0.73$\pm$0.68 \\
Mar 11.908&12.014&11.961&230.973&$-$2.13$\pm$1.51 \\
Mar 14.942&15.126&15.034&92.097&+0.23$\pm$0.34 \\
Mar 14.942&15.126&15.034&219.006&$-$0.25$\pm$1.12 \\ 
\hline
\end{tabular} 
\end{center} 
\end{table} 
\subsection {Optical observations and calibrations}  

The broad band Johnson $UBV$ and Cousins $RI$ observations of the OA were carried
out from 26 to 27 Feb 2003 using 2-m Himalayan Chandra Telescope (HCT) of 
the Indian Astronomical Observatory (IAO), Hanle and the 104-cm Sampurnanand 
telescope of the State Observatory, Nainital. At Nainital, the CCD chip of size 
2048 $\times$ 2048 pixel$^{2}$ covers a field of $\sim 13^{\prime}\times 13^{\prime}$ 
with one pixel corresponding to 0.$^{''}$38 square on the sky. 
The gain and read out noise of the CCD camera are 10 $e^-/ADU$ and 5.3 
$e^-$ respectively.  At Hanle, the CCD used was 1024 $\times$ 1024 pixel$^{2}$,
the entire chip covering a field of $\sim 4^{\prime}.7 \times 4^{\prime}.7$ on 
the sky. It has a read out noise of 11 $e^-$ and gain is 4.8 $e^-/ADU$. During 
good photometric sky conditions, the CCD $BVRI$ observations of the OA field 
along with 
Landolt (1992) standard PG0918+029 region were obtained on 26/27 Feb 2003 for 
calibration purposes. Several twilight flat field and bias frames were also 
obtained during the observing run for the CCD calibrations. 
 
ESO MIDAS, NOAO IRAF and DAOPHOT-II softwares were used for cleaning and image
processing of CCD frames. At Nainital, on the night of 26/27 Feb 2003, atmospheric 
extinction coefficients determined from the observations of PG0918+029 bright stars 
are 0.57, 0.32, 0.20, 0.17 and 0.15 mag in $U, B, V, R$ and $I$ filters respectively. 
The five standard stars in the PG0918+029 region cover a range of $-0.29 < (V-I) < 1.11$
in color and of $12.3 < V < 14.5$ in brightness. They yield the following
transformation equations: 
 
\noindent $u_{CCD} = U - (0.084\pm0.01) (U-B) + (6.83\pm0.01) $  \\ 
            $b_{CCD} = B - (0.062\pm0.01) (B-V) + (4.79\pm0.01) $  \\ 
	    $v_{CCD} = V - (0.034\pm0.01) (B-V) + (4.34\pm0.01) $  \\ 
	    $r_{CCD} = R - (0.034\pm0.01) (V-R) + (4.24\pm0.01) $  \\ 
	    $i_{CCD} = I - (0.048\pm0.01) (V-R) + (4.74\pm0.01) $  \\ 

where $U,B,V,R,I$ are standard magnitudes and $u_{CCD}, v_{CCD}, b_{CCD}, r_{CCD}$ and 
$i_{CCD}$ represent the instrumental magnitudes normalized for 1 second of 
exposure time and corrected for atmospheric extinction. The colour coefficients, 
zero - points and errors in them are determined by 
fitting least square linear regressions to the data points. Using these transformations, 
$BVRI$ photometric magnitudes of 10 secondary standard stars are determined in the GRB 
030226 field and their average values are listed in Table 2. The $(X,Y)$ CCD pixel 
coordinates are converted into $\alpha_{2000}, \delta_{2000}$ values using the 
astrometric positions given by Henden (2003). All these stars, observed 3 to 4 times 
in a filter, have internal photometric accuracy  better than 0.01 mag. 
A comparison between our photometry and that of Henden (2003) yields  
zero-point differences of $0.024\pm0.035, 0.013\pm0.022, 0.018\pm0.021$ and 
$-0.021\pm0.036$ mag in $B, V, R$ and $I$ filters respectively. These numbers are 
based on the comparison of secondary stars having range in brightness from $V = 15$ 
to 17.5 mag.  There is no colour dependence in the photometric differences. We therefore 
conclude that photometric calibration used in this work is secure.

Our observations started about 0.5 day after
the burst and are valuable for dense temporal coverage of the light curve. 
Several short exposures up to a maximum of 30 minutes were generally given
while imaging the OA (see Table 3). In order to improve the signal-to-noise 
ratio of the OA, the data have been binned in $2 \times 2$ pixel$^2$ and also 
several bias corrected and flat-fielded CCD images of OA field taken on a night 
are co-added in the same filter, when found necessary. From these images, 
profile-fitting magnitudes are determined using DAOPHOT-II software. To
determine the difference between aperture and profile fitting magnitudes,
we constructed an aperture growth curve of the well-isolated stars and used
them to determine aperture (about 5 arcsec) magnitudes of the OA. 
They are calibrated differentially with respect to the secondary standards 
listed in Table 2 and the values derived in this way are given in Table 3. 
The distribution of these 34 data points is $N(U,B,V,R,I) = (1,3,7,17,6)$.

The secondary standards are also used to calibrate other photometric 
measurements of OA published in GCN circulars by Ando et al. (2003), Covino et 
al. (2003), Fatkhullin et al. (2003), Garnavich von Braun \& Stanek (2003), Greiner et al. 
(2003b), Guarnieri et al. (2003), Maiorano et al. (2003), Nysewander et al. (2003), 
Price \& Warren (2003), Rumyantsev et al. (2003ab), Semkov (2003) and Von Braun et al. 
(2003). The distribution of these 27 photometric data points is $N(B,V,R,I)
= (2,4,19,2)$.

\begin{table}  
{\bf Table 2.}~The identification number (ID), $(\alpha , \delta)$ for epoch 2000, 
standard $V, (B-V), (V-R)$ and $(V-I)$ photometric magnitudes of the stars in 
the GRB 030226 region are given. Garnavich von Braun \& Stanek (2003) stars
$A$ and $B$ are the stars 6 and 4 respectively. Number of observations taken in 
$B, V, R$ and $I$ filters are 3, 4, 4 and 3 respectively. 
 
\begin{center} 
\scriptsize
\begin{tabular}{ccc cc ccl} \hline  
ID & $\alpha_{2000}$ & $\delta_{2000}$ & $V$& $B-V$ & $V-R$ & $V-I$ \\ 
    & (h m s) & (deg m s) & (mag) & (mag) & (mag) & (mag)  \\ \hline 
 1&11 32 44&25 58 52& 17.01&  0.81&  0.51&  0.96 \\
 2&11 32 46&25 56 56& 16.72&  1.12&  0.64&  1.20 \\
 3&11 32 47&25 51 58& 16.10&  0.62&  0.38&  0.75 \\
 4&11 32 59&25 52 38& 17.45&  0.56&  0.35&  0.68 \\
 5&11 33 00&25 57 13& 15.32&  0.60&  0.36&  0.71 \\
 6&11 33 03&25 51 30& 16.14&  1.16&  0.68&  1.24 \\
 7&11 33 05&25 58 12& 17.29&  0.80&  0.45&  0.88 \\
 8&11 33 18&25 50 30& 15.87&  0.81&  0.48&  0.91 \\
 9&11 33 19&25 48 37& 16.43&  0.66&  0.41&  0.79 \\
 10&11 33 28&25 51 07& 16.46&  1.07&  0.65&  1.22 \\
\hline
\end{tabular} 
\end{center} 
\end{table} 

\normalsize
\begin{table}
 {\bf Table 3.}~CCD $UBVRI$ broad band optical photometric observations of 
the GRB 030226 afterglow. At Hanle, 2-m HCT was used while at Nainital, 104-cm 
Sampurnanand optical telescope was used. 

\begin{center}
\scriptsize
\begin{tabular}{ccll} \hline 
Date (UT) of & Magnitude & Exposure time & Telescope  \\
2003 February& (mag)&(Seconds)&  \\   \hline 
\multicolumn{3}{c}{\bf $U-$ passband}  \\
26.7729&20.17$\pm$0.11&1800 &104-cm \\
\multicolumn{3}{c}{\bf $B-$ passband}  \\
26.6969&20.42$\pm$0.08&2$\times$300 &104-cm \\
26.7326&20.31$\pm$0.07&600 &104-cm \\ 
26.8479&20.82$\pm$0.08&600 &104-cm \\ 
 \multicolumn{3}{c}{\bf $V-$ passband} \\ 
26.6837&19.87$\pm$0.04&2$\times$300& 104-cm   \\ 
26.7335&20.17$\pm$0.03&900&HCT 2-m \\
26.7410&20.25$\pm$0.05&600& 104-cm   \\ 
26.8049&20.31$\pm$0.05&600& 104-cm   \\ 
26.8172&20.33$\pm$0.05&900&HCT 2-m \\
26.8597&20.33$\pm$0.05&900& 104-cm   \\
26.9120&20.68$\pm$0.11&900&HCT 2-m \\ 
 \multicolumn{3}{c}{\bf $R-$ passband} \\ 
26.6398&19.60$\pm$0.06&450&HCT 2-m \\
26.6484&19.64$\pm$0.05&700&HCT 2-m \\
26.6575&19.55$\pm$0.05&700&HCT 2-m \\
26.6674&19.52$\pm$0.04&2$\times$300&104-cm  \\
26.6678&19.62$\pm$0.05&900&HCT 2-m \\ 
26.6805&19.63$\pm$0.04&900&HCT 2-m \\
26.6925&19.67$\pm$0.04&900&HCT 2-m \\
26.7065&19.76$\pm$0.04&900&HCT 2-m \\
26.7216&19.78$\pm$0.04&900&HCT 2-m \\
26.7493&19.74$\pm$0.04&600&104-cm \\ 
26.8139&19.84$\pm$0.04&600&104-cm \\ 
26.8715&20.22$\pm$0.06&900&HCT 2-m \\
26.8715&20.11$\pm$0.04&900&104-cm \\ 
26.8968&20.23$\pm$0.06&900&HCT 2-m \\
26.9932&20.45$\pm$0.06&2$\times$900&HCT 2-m \\
27.8074&21.98$\pm$0.14&900&104-cm \\ 
27.8750&22.00$\pm$0.15&1800&104-cm \\ 
 \multicolumn{3}{c}{\bf $I-$ passband} \\ 
26.6774&19.00$\pm$0.07&2$\times$300&104-cm \\ 
26.7576&19.34$\pm$0.09&600&104-cm \\ 
26.7764&19.59$\pm$0.04&900&HCT 2-m \\
26.8229&19.52$\pm$0.09&600&104-cm \\ 
26.8333&19.51$\pm$0.06&600&HCT 2-m \\
27.8306&21.78$\pm$0.34&2$\times$900&104-cm \\ 
\hline
\end{tabular} 
\end{center} 
\end{table} 

\normalsize
\begin{table}
{\bf Table 4.}~Fitted parameters of GRB 030226 afterglow. The sharpness parameter 
$s$ = 3 is fixed. Large errors in derived parameters of $I$ band may
be due to sparse data-set. 
\begin{center}
\tiny
\begin{tabular}{cccll} \hline 
Fitted & B & V & R & I  \\
Parameters&band&band&band&band  \\   \hline 
$\alpha_1$&.....&.....&0.77$\pm$0.04&..... \\
$\alpha_2$&2.71$\pm$0.17&2.03$\pm$0.07&1.99$\pm$0.06&1.82$\pm$0.34 \\
$t_b$&0.67$\pm$0.06&0.56$\pm$0.07&0.69$\pm$0.04&0.52$\pm$0.32 \\
$m_b$&20.69$\pm$0.16&20.08$\pm$0.19&20.02$\pm$0.06&19.24$\pm$0.95 \\
$\chi^2/DOF$&2.59&3.82&2.06&5.13 \\
\hline
\end{tabular} 
\end{center} 
\end{table}

\section{ Optical photometric light curves}

Fig. 1 shows the light curve of GRB 030226 afterglow in $B, V, R$ and $I$
passbands. Present observations in combinations with above mentioned published 
ones are used in the plot. The X-axis is log $\Delta t (= t-t_0)$. Here $t$ is 
the time of observation and $t_0=$~2003 Feb 26.157315 UT is the burst epoch. 

%----------------------------  FIG. 1  -------------------------------------
\begin{figure}[h]
\centering
\includegraphics[height=11.0cm,width=9.0cm]{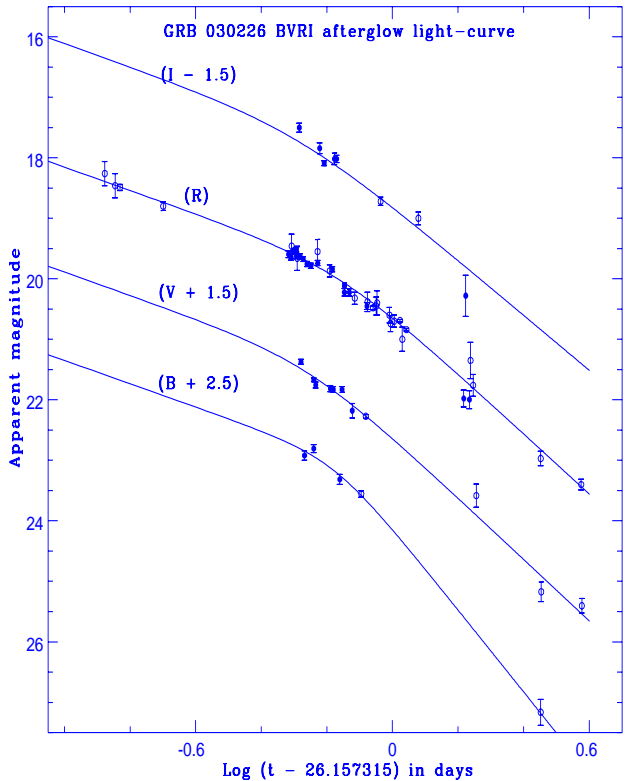}
\caption{\label{light} Optical light curves of GRB 030226 afterglow in  
$BVRI$ passbands. Marked vertical offsets are applied to avoid overlapping 
of data points of different passbands. The present measurements are marked
by solid circles while those taken from the literature are shown by open
circles. The solid curves are the least square best fitted relations for
the parameters listed in Table 4.}
\end{figure}
%----------------------------  FIG. 1  -------------------------------------

The light curve appears to steepen after $\Delta t$ $>$ 0.6 day in $R$
passband. Lack of early observations makes it difficult to see this transition 
in $BVI$ light curves.  To determine the flux decay constants and the break 
time, we fitted the following empirical function (see Rhoads \& Fruchter 2001) 
which represents a broken power-law, to the $R$ band light curve.

\begin{equation}
m=m_b + \frac{2.5}{s} [log_{10} \{ (t/t_b)^{\alpha_1s}
+(t/t_b)^{\alpha_2s}\} - log_{10}(2)] 
\end{equation}

where $\alpha_1$ and $\alpha_2$ are asymptotic power-law slopes at early and 
late times with $\alpha_1 < \alpha_2$ and $s > 0$ controls the sharpness of the 
break, a larger $s$ implying a sharper break. $m_b$ is the magnitude at the 
cross-over time $t_b$. The function describes a light curve falling as 
$t^{-\alpha_1}$ at $t << t_b$ and $t^{-\alpha_2}$ at $t >> t_b$. The results
of the fit are presented in table 4. The value of $s$ has been fixed at $3$,
which helped to achieve the minimum $\chi^2$.
The early time $R$ band data of the OA presented by Ando et al. (2003), 
Price \& Warren (2003) and Von Braun et al. (2003) were 
based on nearby USNO A-2.0 stars in the GRB 030226 field which are brighter by 
0.4--0.5 mag than that determined from our calibration. It is significant to
note that once this difference in calibration is accounted for, there appears
to be no sign of the re-brightening of the OA around 0.5~day after the burst,
contrary to the claim by Dai \& Wu (2003). The $R$ band data can be well 
fitted in terms of broken power law using equation (1). Due to lack of early 
data points in $B,V$ and $I$ passbands, in carrying out fits for these bands
we fixed the value of $\alpha_1$ at 0.77, as determined from $R$-band.
These results are also tabulated in table 4.  The derived value of 
$\alpha_2$ in $B$ band is larger than in other bands, but this may be due to 
sparse late time data points; the deviation is within 3 sigma of the average 
value of deviation in $\alpha_2$. 
In Fig. 1 the best fit light curves obtained in this way for $BVRI$ passbands 
are shown. It can be seen that the observations presented here 
lie near the jet break time, filling important gaps in the published data.

In the light of the above, we conclude that the early time light curve decay
slope $\alpha_1$ is 0.77$\pm$0.04 and the weighted mean values of 
parameters $t_b$ and $\alpha_2$ are 0.65$\pm$0.03 and 2.05$\pm$0.04
respectively. The value of $\alpha_2$ is flatter than the observed 
$X-$ ray flux decay of 3.6$\pm$1.0 (Sako \& Fox 2003) but agrees well with 
$\alpha_2$ $\sim$ 2 determined by Klose et al. (2003) and Dai \& Wu (2003). 
The value 
of $s = 3$ indicates that the observed break in the light curve is sharp, 
unlike the smooth break observed in the optical light curve of GRB 990510 
(Stanek et al. 1999; Harrison et al. 1999) 
but similar to the sharp break observed in the optical light curves of GRB 000301c
(Sagar et al. 2000, Pandey et al. 2001); GRB 000926 (Harrison et al. 2001, Sagar 
et al. 2001a, Pandey et al. 2001); GRB 010222 (Masetti et al. 2001, Sagar et al. 
2001b, Stanek et al. 2001, Cowsik et al. 2001) and GRB 011211 (Jakobsson et al. 2003). 

%----------------------------  FIG. 2  --------------------------------------
\begin{figure}[h]\centering
\includegraphics[height=11.0cm,width=9.0cm]{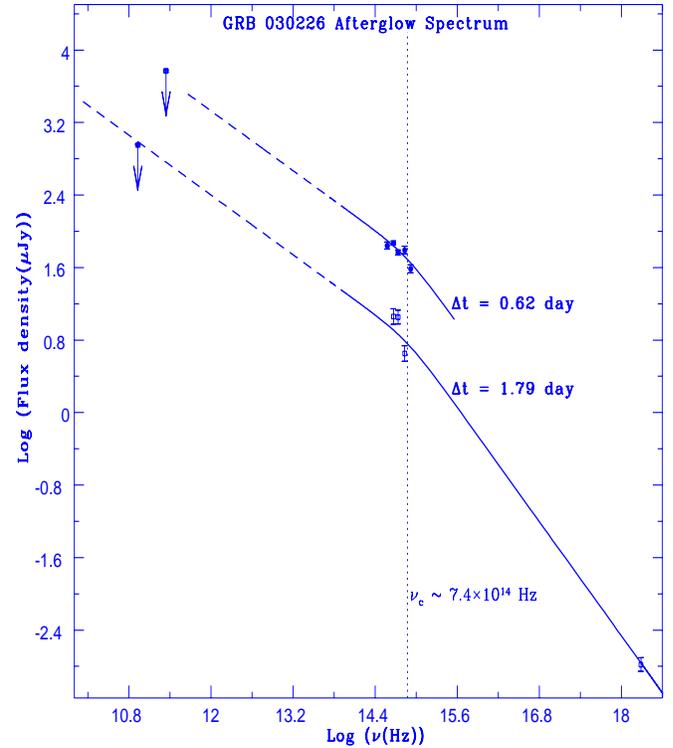}
\caption{\label{light} Spectral energy distribution (SED) of GRB 030226 
afterglow from millimeter to X-ray passbands are shown at two marked epochs. 
The optical data are corrected for a small Galactic extinction of $A_V$ = 0.06 
mag and a starburst galaxy-like extinction (Calzetti et al. 1997) in the host 
galaxy, of amount $E(B - V) = 0.17$ mag. Arrows show the millimeter upper limits
(85 GHz and 220 GHz) at $\Delta t$ = 1.79 day. Filled and open circles show
the data at $\Delta t$ = 0.62 and 1.79 day respectively. The location of 
the cooling frequency $\nu_c$ is shown by a dotted line.}
\end{figure}
%----------------------------  FIG. 2  --------------------------------------

\section{Spectral Energy Distribution}

Fig. 2 shows the spectrum of GRB 030226 afterglow from millimeter to $X-$ ray region.
We construct the GRB 030226 afterglow spectrum at two possible epochs to cover
the longest spectral range: $\Delta t$ = 0.62 day, an epoch near the jet break,
and $\Delta t$ = 1.79 day, corresponding to the epoch of $X-$ ray 
observations. We also estimate the millimeter upper limits from the present 
measurements for the epoch $\Delta t$ = 1.79 day. The reddening map provided by 
Schlegel, Finkbeiner \& Davis (1998) indicates a small value of $E(B - V) = 0.02$ 
mag for the Galactic interstellar extinction towards the burst. We used the standard 
Galactic extinction reddening curve given by Mathis (1990) to convert apparent 
magnitudes into fluxes, with the effective wavelengths and normalizations from 
Fukugita et al. (1995) for $UBVRI$. $X-$ ray measurements were obtained by Pedersen 
et al. (2003) in 2 -- 10 KeV energy range. Considering mid UT as the epoch and 20 \% 
uncertainty in the measurements, we calculate the $X-$ ray flux at $\Delta t$ = 1.79 
day. For the epochs under study it is observed that the flux increases as the 
frequency decreases. 

In the simple synchrotron model of the GRB afterglow, the light curves and spectral 
energy distributions are generally described as a power law: $F(t, \nu) 
\propto t^{-\alpha} \nu^{-\beta}$.  The spectral slope $\beta$ depends on the 
location of spectral breaks. So does the decay slope $\alpha$, which also has 
an additional dependence on the dynamics of the fireball. However
within each regime $\alpha$ and $\beta$ are functions only of $p$, the power 
law exponent of the electron Lorentz factor. For an initially collimated ejection, 
at late times when the evolution comes to be dominated by the lateral spreading of the 
jet, the value of $\alpha$ is expected to approach $p$. The value of $\beta$ is expected 
to be $p/2$ above the cooling frequency $\nu_c$ and $(p - 1)/2$ below it, down to
the peak of the spectrum, in the slow cooling case (Sari, Piran \& Halpern 1999).
In the present case the post jet-break flux decay constant is determined to be
$\alpha_2 = 2.05\pm0.04$, indicating that the electron energy distribution index
$p$ must be close to this value (Rhoads 1999).
At $\Delta t = 0.62$~day, the fitted spectrum through the optical band has 
spectral slope $\beta \sim 1.0$ and at $\Delta t = 1.79$~day, $\beta \sim 1.2$. The 
determined values of $\beta$ at both epochs are similar and are consistent with
the measured $\alpha_2 = 2.05\pm0.04$ if $\nu > \nu_c$ in the optical band.
However this predicts a pre jet-break decay constant $\alpha_1$ around 1.1, 
much steeper than the observed value of $0.77\pm0.04$. So the cooling frequency
is unlikely to be below the optical band.  A consistent picture, including the
measured X-ray flux, can be obtained if the cooling frequency is located just 
above the optical band, and the intrinsic optical spectrum of the afterglow
is flatter, with a slope $\sim 0.55$.  Extinction along the line of sight would
steepen the optical spectrum, and could explain the observed $\beta \sim 1$.
The estimated Galactic extinction of $A_V = 0.06$~mag does not change the slope
much, so we attribute the rest of the extinction to the host galaxy of the GRB.
Assuming $p=2.1$, we find that $E(B - V) \approx 0.15$ mag in the host galaxy, 
with an extinction law similar to that in starburst galaxies (Calzetti et al.\ 
1997), can reproduce the expected spectrum. Assuming that the cooling
frequency $\nu_c$ is located $\sim 7\times10^{14}$~Hz, the result is consistent 
also with the X-ray flux observed at $1.79$~day (see Fig.~2).
The cooling frequency is expected to be nearly the same even in the 
$0.62$~day spectrum since both epochs fall after the jet break.
We found that a host extinction law similar to that for our galaxy 
(Cardelli et al.\ 1989) is unable to provide a consistent fit: the presence 
of the 2200\AA\ bump introduces additional scatter in the data.  
A combined fit to spectra at both epochs, with $p$ and $\nu_c$ fixed to
the above values, yields the following estimate for the extinction in the host
galaxy: $E(B-V) \sim 0.17$ mag.

The two millimeter upper limits at $\Delta t = 1.79$ day lie close to or
above the extrapolation of the optical spectrum, and are therefore consistent
with the above interpretation.  It may be noted that the above derived
parameters have an accuracy of $\sim$ 10 -- 20 \% only, as they are based
on a few available observed data points.

\section{Discussion and Conclusions}

We have presented optical photometry of the afterglow of GRB~030226 in $UBVRI$ 
bands from 0.5 to 2 days and millimeter observations for up to 20 days after 
the burst. We do not detect millimeter wave emission down to a few mJy.
The overall flux decay observed in $R$ band is well understood in terms of a 
jet model.  The flux decay constants at early and late times are 0.77$\pm$0.04 
and 2.05$\pm$0.04 respectively, and the jet break time is 0.65$\pm$0.05 day. 
These decay indices indicate a value of the electron energy distribution
index $p=2.07\pm0.06$.  Comparison of the observed broadband spectrum with
that predicted by synchrotron emission model suggests an extinction
$E(B - V) \sim 0.17$ mag in the host galaxy, and an extinction law similar 
to that found in starburst galaxies (Calzetti et al.\ 1997).  The cooling break 
is inferred to be located at the upper edge of the optical band,
$\nu_c \sim 7\times 10^{14}$~Hz.

GRB~030226 belongs to a small subset of GRBs which display a steepening in the 
optical afterglow light curve (``jet break'') within a day after the burst.  
Other known cases include GRB~980519 ($\sim 0.6$~day, Jaunsen et al.\ 2001), 
GRB~010222 ($\sim 0.7$~day, Stanek et al.\ 2001, Sagar et al.\ 2001b), 
GRB~020813 ($\sim 0.2$~day, Bloom et al.\ 2002) and GRB~030329 
($\sim 0.6$~day, Burenin et al.\ 2003). 
Of these, GRB~980519 showed a cooling break between optical
and X-ray bands (Jaunsen et al.\ 2001), GRB~010222 had its cooling break
below the optical band (Sagar et al.\ 2001b) and in the present case we
infer the cooling break to be at the upper edge of the optical band.
GRB~010222 also exhibited a hard electron energy spectrum $p<2$
(Bhattacharya 2001, Sagar et al.\ 2001b).

The observed fluence of 5.7 $\mu$erg/cm$^2$ in the energy band 30 -- 400 KeV 
with the measured redshift z = 1.986$\pm$0.001 indicates an isotropic 
equivalent energy release $E_{\rm iso}\sim 6.5 \times 10^{52}$~erg for 
$H_0$ = 65 km/s/Mpc in a $\Omega_0$ = 0.3 and $\Lambda_0$ = 0.7 cosmological 
model. The observed jet-break time of $0.6$~day leads to an estimated
jet half-opening angle of $\sim 3.2$ degree, for an assumed particle density
n = 1~cm$^{-3}$. The total energy output in the jet then works out to be
$\sim 0.8\times 10^{51}$~erg for a $\gamma-$ray efficiency $\eta_{\gamma}=0.2$,
after applying the cosmological K-correction (Bloom et al.\ 2001). 
This is close to the estimated mean energy output in Gamma Ray Bursts
(Frail et al.\ 2001, Kulkarni et al.\ 2003, Berger et al. 2003), and 
supports the case for GRBs as standard energy reservoirs.

\section*{Acknowledgements}
This research has made use of data obtained through the High Energy 
Astrophysics Science Archive Research Center Online Service, provided by the 
NASA/Goddard Space Flight Center. Thanks to anonymous referee for the
useful comments.


\begin{thebibliography}{}

\bibitem {} Ando, M., Ohta, K., Watanabe, C. et al. 2003, GCNC 1882, 1884
\bibitem {} Berger, E., Kulkarni, S.R., Pooley, G., et al., 2003,
	    Nature, in press (astro-ph/0308187)
\bibitem {} Bhattacharya D., 2001, BASI, 29, 107
\bibitem {} Frail, D. A., Kulkarni, S. R., Sari R. et al., 2001, ApJ, 562, L55
\bibitem {} Bloom, J. S., Frail, D. A. \& Sari, R. 2001, AJ, 121, 2879 
\bibitem {} Bloom, J. S., Fox, D. W. \& Hunt, M. P. 2002, GCNC 1476 
\bibitem {} Burenin, R. A., Sunyaev, R. A., Pavlinsky, M. N. et al., 2003, Astr. Lett. 29, 9
\bibitem {} Calzetti, D. 1997, AJ, 113, 162
\bibitem {} Cardelli, J. A., Clayton, G. C. \& Mathis, J. S. 1989, ApJ, 345, 245
\bibitem {} Cowsik, R., Prabhu, T. P., Anupama, G. C. et al. 2001, BASI, 29, 157
\bibitem {} Chornock, R. \& Fillippenko, A. V. 2003, GCNC 1897
\bibitem {} Covino, S., Ghisellini, G., Malesani, D. et al. 2003, GCNC 1909
\bibitem {} Dai Z. G. \& Wu X. F., 2003, ApJL, 591, L21 
\bibitem {} Fatkhullin, T., Komarova, V., Sokolov, V. Cherepashchuk, A., \& Postnov. K. 2003, GCNC 1925
\bibitem {} Fox, D. W., Chen, H. W. \& Price, P. A. 2003, GCNC 1879 
\bibitem {} Fukugita, M., Shimasaku, K. \& Ichikawa, T. 1995, PASP, 107, 945 
\bibitem {} Garnavich, P., Von Braun, K. \& Stanek, K. 2003, GCNC 1885
\bibitem {} Greiner, J., Guenther, E., Klose., S. \& Schwarz., R. 2003a, GCNC 1886
\bibitem {} Greiner, J., Ries, C, Barwig, H., Fynbo, J. \&  Klose, S. 2003b, GCNC 1894
\bibitem {} Guarnieri, A., Cortese, L., Bartolini, C. et al. 2003, GCNC 1892
\bibitem {} Guilloteau, S., Delannoy, J., Downes, D. et al. 1992, A\&A, 262, 624
\bibitem {} Harrison, F. A., Bloom, J. S., Frail, D. A. et al. 1999, ApJ, 523, L121
\bibitem {} Harrison, F. A., Yost, S. A., Sari, R. et al. 2001, 559, 123
\bibitem {} Henden A., 2003, GCNC 1916
\bibitem {} Jakobsson, P., Hjorth, J., Fynbo, J. U. et al. 2003, Accepted in A\&A
\bibitem {} Jaunsen, A. O., Hjorth., J., Bj\"{o}rnsson, G. et al. 2001, ApJ, 546, 127
\bibitem {} Klose, S., Stecklum, B., Zeh, A. et al. 2003, GCNC 1923
\bibitem {} Kulkarni, S. R., Fox, D. W., Berger, E. \& Soderberg A. M. et al. 2003, GCNC 1911
\bibitem {} Landolt, A.R., 1992, AJ, 104, 340 
\bibitem {} Maiorano, E., Masetti, N., Palazzi, E. et al. 2003, GCNC 1933
\bibitem {} Masetti, N., Palazzi, E., Pian, E. et al. 2001, A\&A, 374, 382 
\bibitem {} Mathis J.S., 1990, ARAA, 28, 37 
\bibitem {} M\'{e}sz\'{a}ros P., 2002, ARA\&A, 40, 137 
\bibitem {} Nysewander, M. C., Moran, J., Reichart, D. \& Schwartz, M. 2003, GCNC 1921
\bibitem {} Pandey, S.B., Sagar, R., Mohan, V. et al. 2001, BASI, 29, 459 
\bibitem {} Pedersen, K., Fynbo, J., Hjorth, J. \& Watson, D. et al. 2003, GCNC 1924
\bibitem {} Price P. A. \& Warren B. E., 2003, GCNC 1890 
\bibitem {} Price P. A., Fox D. W. \& Chen H. W., 2003, GCNC 1880
\bibitem {} Price, P. A., Fox D. W., Djorgovski, S. G. et al. 2003, GCNC 1889
\bibitem {} Rhoads J.E., 1999, ApJ, 525, 737 
\bibitem {} Rhoads J. E. \& Fruchter A., 2001, ApJ, 546, 117
\bibitem {} Rumyantsev, V., Biryukov, V. \& Pozanenko A. 2003a, GCNC 1908 
\bibitem {} Rumyantsev, V., Sergeeva, L. \& Pozanenko A. 2003b, GCNC 1929 
\bibitem {} Sako M \& Fox D. W., 2003, GCNC 1928
\bibitem {} Sagar, R., 2002, BASI, 30, 237
\bibitem {} Sagar, R., Mohan, V., Pandey, S.B. et al. 2000, BASI, 28, 499 
\bibitem {} Sagar, R., Pandey, S.B., Mohan, V., Bhattacharya, D. \& Castro-Tirado, A.J. 2001a, BASI, 29, 1,
\bibitem {} Sagar, R., Stalin, C. S., Bhattacharya, D. et al. 2001b, BASI, 29, 91
\bibitem {} Sari, R., Piran, T., Halpern, J. P. 1999, ApJ, 519, L17 
\bibitem {} Schlegel, D.J., Finkbeiner, D.P., Davis, M. 1998, ApJ, 500, 525 
\bibitem {} Schwarz, P. R. 1980, PASP, 535, 534
\bibitem {} Semkov, E. 2003, GCNC 1935
\bibitem {} Stanek, K. Z., Garnavich, P. M., Kaluzny, J., Pych, W. \& Thompson, I. 1999, ApJ, 522, L39 
\bibitem {} Stanek, K. Z., Krzysztof, Z., Garnavich, P. M. et al. 2001, ApJ, 563, 592 
\bibitem {} Suzuki, M., Shirasaki, Y., Graziani, C. et al., 2003, GCNC 1888
\bibitem {} Von Braun, K., Garnavich, P. \& Stanek, K.  2003, GCNC 1881
\end{thebibliography}
\end{document}